\documentstyle[amssymb,12pt]{article}

\begin{document}

\title{Dark matter, dark charge, and the fractal structure of the Universe }
\author{A.A. Kirillov \\
{\em Institute for Applied Mathematics and Cybernetics} \\
{\em 10 Uljanova Str., Nizhny Novgorod, 603005, Russia} \\
e-mail: kirillov@unn.ac.ru}
\date{}
\maketitle

\begin{abstract}
It is shown that the observed fractal distribution of galaxies is, in fact,
consistent with homogeneity of the Universe and observational limits on $ 
\Delta T/T$, if the presence of dark matter and dark charge predicted by the
Modified Field Theory is taken into account. It is discussed a new scenario
of structure formation in which observed structures appear as a result of
decay of the primordial fractal distribution of baryons.
\end{abstract}

As it was shown (e.g., see Ref. \cite{P87,CP} and for a more recent
discussions Ref. \cite{LMP,R99,FrD}) observed galaxy distribution exhibits a
fractal behaviour with dimension $D\approx 2$ and which seems to show no
evidence of cross-over to homogeneity. In spite the fact that the picture in
which baryons have a fractal distribution in space is not widely accepted,
it, in fact, has recently found a rigorous theoretical ground. First, we
note that there are two basic arguments which do not allow to accept such a
picture. The first argument is that the fractal distribution can be in a
conflict with the Cosmological Principle which is expressed by the statement
that the metric of the universe is well approximated by the Friedmann metric
and, therefore, gravitational potential fluctuations must be small. Such
metric potential fluctuations are measured directly by $\Delta T/T$ in the
microwave background and as it was discussed in Ref. \cite{Bar} the density
distribution does not give direct evidence of the ''Cosmological
principle''. However, if we suppose that the Universe does not contain some
hidden component of matter and all baryons are described by a fractal
distribution in space, this should result in too strong metric fluctuations
and too strong $\Delta T/T$, whereas there exist strong observational limits 
$\Delta T/T<10^{-5}$. We stress that if the leading contribution to the
matter density is given by a dark component then metric fluctuations are not
directly related to the distribution of baryons and can remain small.
However, there exists another argument which does not depend on the presence
of dark matter and can be considered independently. It is the Silk effect
which directly relates the distribution of baryons at the moment of
recombination with variation of the CMB temperature $n_{b}\sim T^{3}$ and,
therefore, the fractal distribution of baryons must leave a direct imprint
in the temperature. It turns out that these problems can be easily solved in
the so-called Modified Field Theory (MOFT) suggested first in Ref.\cite{k99}
and developed recently in Ref.\cite{KT}.

Indeed, MOFT was shown to possess a nontrivial vacuum state which leads to a
scale-dependent renormalization of all interaction constants $\alpha
^{2}\rightarrow $ $\alpha ^{2}N\left( k\right) $ (where $k$ is the wave
number and $\alpha $ is either the electron charge $e$, the gauge charge $g$
, or the gravitational constant $\sqrt{G}$) with the same structural
function $N\left( k\right) $ which reflects the topology of the momentum
space (for details we send readers to Ref. \cite{KT}). This means that in
MOFT particles lose their point-like character and acquire a specific
distribution in space, i.e., each point source is surrounded with a dark
halo which carries charges of all sorts. In the case of the
thermodynamically equilibrium state at low temperatures properties of the
dark halo are described by two characteristic scales $\ell _{\min }$ and $ 
\ell _{\max }$ between which interaction constants do increase $\alpha
^{2}\left( \ell \right) \sim \alpha _{0}^{2}\ell /\ell _{\min }$ ($\ell
=2\pi /k$) and the Newton's and Coulomb's interaction energies show the
logarithmic $V\sim \ln r$ (instead of $1/r$) behaviour. We note that MOFT
reproduces the basic features of the Modified Newtonian dynamics proposed by
Milgrom Ref. \cite{mil} (see also the list of references devoted to MOND at
the site \cite{Site}). However conceptually MOFT differs form the latter
and, besides, predicts analogous modification of the Maxwell's
electrodynamics.

Consider a particular baryon as a source for electromagnetic and
gravitational field and a test particle. Due to the halo both particles are
distributed in space and the energy of interaction between them (in the case
of rest particles) in the range $\ell _{\min }$ $<r<$ $\ell _{\max }$\ is
(see Ref.\cite{KT}) 
\begin{equation}
V\left( r\right) \sim \frac{q_{b}q}{\ell _{\min }}\ln \left( \frac{r}{\ell
_{\max }}\right) ,  \label{pot}
\end{equation}
where $q_{b}q=m_{p}mG$ ($m_{p}$ is the proton mass, $m$ \ is the mass of the
test particle, and $G$ is the gravitational constant) for the gravitational
interaction and $q_{b}q=-Ze^{2}$ (where $Ze$ is the charge of the test
particle) for the electromagnetic field. If we suppose the law $1/r$ is
intact and the test particle has a point-like character, we define the
''dynamical'' charge/mass of the baryon contained within a radius $r$ with
the baryon in the center as $q\left( r\right) \sim q_{b}r/\ell _{\min }\ln
\left( r/\ell _{\max }\right) .$ Consider a baryon only Universe, then the
total dynamical charge and \ the dynamical mass within the radius $r$ will
be given by 
\begin{equation}
Q_{tot}\left( r\right) =q\left( r\right) N_{b}\left( r\right) +\delta
Q\left( r\right) ,  \label{ch}
\end{equation}
where $N_{b}\left( r\right) $ is the actual number of baryons and $\delta
Q\left( r\right) $ accounts for the contribution of dark halos of baryons \
from the outer region.

It is important that for potentials of the type (\ref{pot}) the homogeneous
distribution of particles (which corresponds to $N\left( r\right) \sim
nr^{3} $ with a constant $n$) is unstable (at least gravitationally). This
can be seen straightforwardly from ( \ref{pot}) by considering the limit $ 
\ell _{\max }\rightarrow \infty $, or $\ell _{\min }\rightarrow 0$. In this
limit the dynamical charge diverges $q\left( r\right) \rightarrow \infty $,
the interaction energy of particles becomes too strong, and such a system
collapses. In this case a stable distribution of particles can be reached by
a fractal distribution $N\left( r\right) \sim r^{D}$ with $D\approx 2$.
Indeed, let us consider the ''dynamical'' (or effective) number of particles 
$\widetilde{N}_{b}=Q_{tot}\left( r\right) /q_{b}$. Then the equilibrium
state will corresponds to the homogeneous distribution of the dynamical (or
effective) density of particles, i.e., $\widetilde{n}\sim \widetilde{N}
_{b}/r^{3}=const$. This, in turn is consistent with both, the Cosmological
Principle (Friedmann metric) and the Silk effect ($T^{3}\sim \widetilde{n}
=const$), while the actual number of baryons in the range $\ell _{\min
}\lesssim r\lesssim $ $\ell _{\max }$, as it can be seen from (\ref{ch}),
will follow the law $N_{b}\left( r\right) \leq Q_{tot}\left( r\right)
/q\left( r\right) $ $\sim $ $\widetilde{n} r^{2}\ell _{\min }/\ln \left(
r/\ell _{\max }\right) $ which perfectly fits the results of Ref. \cite{LMP} 
. Above the scale $\ell _{\max }$ and below $\ell _{\min }$ the\ law $1/r$
restores and the actual distribution of baryons has to cross-over to
homogeneity $N_{b}\left( r\right) \sim r^{3}$. The minimal scale $\ell
_{\max }$ represents the lower boundary where effects of MOFT start to show
up and as it is shown in Ref.\cite{kin} $\ell _{\min }\sim 4\div 5kpc$. The
results of Ref. \cite{LMP} provide that the size of the upper cutoff $\ell
_{\max }$ , if it really does exist, must be more than $200Mpc$ and,
therefore, for the baryon fraction we get the limit $\Omega _{b}/\Omega
_{tot}\sim \ell _{\min }/\ell _{\max }\lesssim 10^{-5}$ (which, in
particular, sets the restriction on the mass of the primordial scalar field $ 
m<10^{-5}m_{pl}$). We recall that, in fact, due to the dark charge the Silk
effect does not set any restrictions on possible values of $\ell _{\max }$
and it can be larger than the Hubble radius.

To clarify the reason of origin of the fractal distribution of baryons we
note that as it was shown in Ref. \cite{k99} the nontrivial topology of the
momentum space ($N\left( k\right) \neq 1$) leads to the fact that the ground
state of all bosonic fields is characterized by a nonvanishing particle
number density. Consider plasma oscillations of baryons. These oscillations
represent Bose particles and in MOFT the total energy of such oscillations
can be written as $H=\sum_{k}n\omega _{k}N\left( n,k\right) $, where $ 
N\left( n,k\right) $ is the number of field (plasma) modes with the given
wave number $k$ and the number of plasmons $n$, while $\omega _{k}$ is the
energy of a plasmon (for small $k$ we get $\omega _{k}$ $=$ $\omega _{0}$ $=$
$\left( 4\pi e^{2}\widetilde{n}/m_{p}\right) ^{1/2}$ where $\widetilde{n}$
is the dynamical density introduced above which accounts for the
renormalization of the charge). The ground state is characterized by
occupation numbers $N\left( n,k\right) =\theta \left( \mu _{k}-n\omega
_{k}\right) $ where $\theta \left( x\right) $ is the step function and the
chemical potential $\mu _{k}$ is defined by $\sum_{n}N\left( n,k\right) $ $= 
$ $1+\left[ \mu _{k}/\omega _{k}\right] $ $=$ $N\left( k\right) $, where $ 
[x] $ denotes the integral part of $x$ and $N\left( k\right) $ is the
structural function which defines the topology of the momentum space (see,
Ref. \cite{KT}). In the range $\ell _{\max }\geq 2\pi /k$ $\gg \ell _{\min }$
the function $N\left( k\right) $ can be approximated by $N_{k}\sim 1+\left[
k_{0}/k\right] $ ($k_{0}=2\pi /\ell _{\min }$). The number of plasmons in
the ground state is $\overline{n}_{k}=\sum_{n}nN\left( n,k\right) =\frac{1}{ 
2 }N\left( k\right) \left( N\left( k\right) -1\right) $, i.e., is solely
expressed via the structural function, and thus, the energy of the ground
state reads $H_{0}\simeq \sum_{k}\overline{n}_{k}\omega _{0}$ (for the sake
of simplicity we suppose $\omega _{0}\gg 2\pi c/\ell _{\min }$). From the
other hand side this energy must coincide with the energy of baryons which
at low temperatures is given by $H_{0}\simeq \sum_{k}m_{p}n_{bk}$. Thus, we
see that the baryon number density should follow the density of plasmons in
the ground state $n_{b}\simeq $ $\overline{n}_{k}\omega _{0}/m_{p}$ which at
large scales behaves as $\overline{n}_{k}\sim $ $\left( k_{0}/k\right) ^{2}$
, e.g., corresponds to the fractal distribution.

We stress that in MOFT observed fractal galaxy distribution (at large
scales) corresponds to the thermodynamically equilibrium state at low
temperatures and in this sense is not a result of an evolutionary process as
it is commonly considered. The fractal distribution of baryons represents a
background which corresponds to the homogeneous distribution of matter,
while gravitational potential fluctuations can be considered in the same way
as in the standard model which, however, to make specific predictions,
requires the further investigation.

In conclusion we note that there is a possibility to construct a new
scenario in which the observed structures (stars, galaxies, clusters, etc.)
is the result of a decay of the primordial equilibrium fractal distribution
of baryons. Indeed, the only relevant parameter in MOFT is the minimal scale 
$\ell _{\min }$ below which there acts the standard Newton's physics and the
fractal distribution of baryons is unstable. If, at present, topology
changes of the momentum space are forbidden, then this scale must be a
constant in the commoving frame $\ell _{\min }/a\left( t\right) =\widetilde{
\ell }_{\min }=const$ (where $a\left( t\right) $ is the scale factor). If
this is the case, then observed structures can appear only due to the growth
of primordial density fluctuations (i.e., in the same way as in the standard
model). However, the primordial topology of the momentum space may
eventually decay and this would lead to a monotonic increase of the minimal
scale $\widetilde{\ell }_{\min }\left( t\right) $. Then on early stages the
minimal scale could reach the value $\ell _{\min }\sim r_{0}$ (where $r_{0}$
is a characteristic baryon size) and the primordial distribution of baryons
followed the fractal behaviour at all scales ($n_{b0}\sim b/k^{2}$).
Eventually $\widetilde{\ell }_{\min }$ increases and due to the instability
of the fractal baryon distribution, which takes place below the scale $ 
\widetilde{\ell }_{\min }$, the larger and larger scales are involved into
the structure formation.

The simplest realization of the fractal law $N_{b}\left( r\right) \sim r^{2}$
gives the case when baryons are homogeneously distributed on thin two-
dimensional surfaces and at present $\ell _{\max }\sim 4\div 5kpc$. This may
give at least a qualitative explanation of why at scales $\ell >\ell _{\min
} $ we observe spiral galaxies (which have the size $L\sim 10\div 20kpc$ $ 
>\ell _{\min }$) which are still flat, while at smaller scales $\ell <\ell
_{\max }$ we observe only spherical collections of stars ($L\sim 1pc<\ell
_{\min }$).

I would like to acknowledge the referee for valuable comments which
essentially improved this Letter.


\begin{thebibliography}{99}
\bibitem{P87}  Pietronero L., 1987, Physica, A144, 257

\bibitem{CP}  Coleman P.H., Pietronero L., Phys. Rep., {\bf 231}, 311 (1992).

\bibitem{LMP}  S. F. Labini, M. Montuori, L. Pietronero: Phys. Rep.{\bf 293}
, 66 (1998).

\bibitem{R99}  K.K.S.Wu, O.Lahav, M. Rees, Nature {\bf 397} 225 (1999).

\bibitem{FrD}  Pietronero L., ``The Fractal debate''; http: //www. phys.
uniroma1. it/ DOCS/ PIL/ pil.html

\bibitem{Bar}  J.D. Barrow, QJRAS 30, 163 (1989).

\bibitem{k99}  A.A. Kirillov, JETP {\bf 88} 1051 (1999) [Zh. Eksp. Teor.
Fiz. {\bf 115} 1921 (1999)], hep-th/9911168.

\bibitem{KT}  A.A. Kirillov, D. Turaev, Phys. Lett. {\bf B 532} 185 (2002),
astro-ph/0202302.

\bibitem{mil}  M. Milgrom, ApJ, {\bf 270} 365 (1983); ApJ, {\bf 270} 371
(1983); ApJ, {\bf 270} 384 (1983).

\bibitem{Site}  http://www.astro.umd.edu/\symbol{126}ssm/mond/litsub.html

\bibitem{kin}  W.H. Kinney, M. Brisudova, Proc. of 15th Florida Workshop in
Nonlinear Astronomy and Physics, ''The Onset of Nonlinearity'',
astro-ph/0006453 (2000).
\end{thebibliography}
\end{document}